\documentclass[twocolumn]{aastex6}
\begin{document}


\title{Clustering of infrared-bright dust-obscured galaxies revealed by the Hyper Suprime-Cam and WISE}

%% Use \author, \affil, plus the \and command to format author and affiliation 
%% information.  If done correctly the peer review system will be able to
%% automatically put the author and affiliation information from the manuscript
%% and save the corresponding author the trouble of entering it by hand.
%%
%% The \affil should be used to document primary affiliations and the
%% \altaffil should be used for secondary affiliations, titles, or email.

%% Authors with the same affiliation can be grouped in a single

\author{Yoshiki Toba 		\altaffilmark{1,2},
		Tohru Nagao 		\altaffilmark{2},
		Masaru Kajisawa		\altaffilmark{2,3},
   		Taira Oogi	    	\altaffilmark{4},   
   		Masayuki Akiyama	\altaffilmark{5},
   		Hiroyuki Ikeda		\altaffilmark{6},  
   		Jean Coupon			\altaffilmark{7},
   		Michael A. Strauss	\altaffilmark{8},
   		Wei-Hao Wang		\altaffilmark{1},  
   		Masayuki Tanaka		\altaffilmark{6}, 
		Mana Niida			\altaffilmark{3},		 	
		Masatoshi Imanishi	\altaffilmark{6,9},
		Chien-Hsiu Lee	    \altaffilmark{6,10}, 
		Hideo Matsuhara 	\altaffilmark{11,12},
		Yoshiki Matsuoka	\altaffilmark{6,9}, 
		Masafusa Onoue		\altaffilmark{6,9},
		Yuichi Terashima	\altaffilmark{2,3},
		Yoshihiro Ueda		\altaffilmark{13},  
		Yuichi Harikane		\altaffilmark{14,15},
		Yutaka Komiyama 	\altaffilmark{6,9},
		Satoshi Miyazaki	\altaffilmark{6,9}, 
		Akatoki Noboriguchi	\altaffilmark{3},	
		Tomonori Usuda		\altaffilmark{6,9}
		}	  
\affil{}  
  \altaffiltext{1}{Academia Sinica Institute of Astronomy and Astrophysics, PO Box 23-141, Taipei 10617, Taiwan}
  \email{toba@asiaa.sinica.edu.tw}
  \altaffiltext{2}{Research Center for Space and Cosmic Evolution, Ehime University, Bunkyo-cho, Matsuyama, Ehime 790-8577, Japan} 
  \altaffiltext{3}{Graduate School of Science and Engineering, Ehime University, Bunkyo-cho, Matsuyama 790-8577, Japan}
  \altaffiltext{4}{Kavli Institute for the Physics and Mathematics of the Universe (Kavli IPMU, WPI), The University of Tokyo, 5-1-5 Kashiwanoha, Kashiwa, Chiba 277-8583, Japan}   
  \altaffiltext{5}{Astronomical Institute, Tohoku University, Aramaki, Aoba-ku, Sendai, 980-8578, Japan}
  \altaffiltext{6}{National Astronomical Observatory of Japan, 2-21-1 Osawa, Mitaka, Tokyo 181-8588, Japan}
  \altaffiltext{7}{Astronomical Observatory of the University of Geneva, ch. d'Ecogia 16, CH-1290 Versoix, Switzerland}
  \altaffiltext{8}{Princeton University Observatory, Peyton Hall, Princeton, NJ 08544, USA}
  \altaffiltext{9}{Department of Astronomy, School of Science, SOKENDAI (The Graduate University for Advanced Studies), 2-21-1 Osawa, Mitaka, Tokyo 181-8588, Japan}
  \altaffiltext{10}{Subaru Telescope, 650 North A'ohoku Place, Hilo, HI 96720, USA}
  \altaffiltext{11}{Institute of Space and Astronautical Science, Japan Aerospace Exploration Agency, 3-1-1 Yoshinodai,Chuo-ku, Sagamihara, Kanagawa 252-5210, Japan}
  \altaffiltext{12}{Department of Space and Astronautical Science, SOKENDAI (The Graduate University for Advanced Studies), 3-1-1 Yoshinodai, Chuo-ku, Sagamihara, Kanagawa 252-5210, Japan}
  \altaffiltext{13}{Department of Astronomy, Graduate School of Science, Kyoto University, Kitashirakawa-Oiwake cho,Sakyo-ku, Kyoto, Kyoto 606-8502, Japan}
  \altaffiltext{14}{Institute for Cosmic Ray Research, The University of Tokyo, 5-1-5 Kashiwanoha, Kashiwa, Chiba 277-8582, Japan}
  \altaffiltext{15}{Department of Physics, Graduate School of Science, The University of Tokyo, 7-3-1 Hongo, Bunkyo, Tokyo, 113-0033, Japan} 
 
%\author{Butler Burton\altaffilmark{3}}
%\affil{National Radio Astronomy Observatory}
%\affil{TeXnology Inc}

%\author{Julie Steffen\altaffilmark{4}}
%\affil{American Astronomical Society \\
%2000 Florida Ave., NW, Suite 300 \\
%Washington, DC 20009-1231, USA}

%% Use the \and command so offset the last author.
%\and

%\author{Jeff Lewandowski\altaffilmark{5}}
%\affil{IOP Publishing, Washington, DC 20005}


%\altaffiltext{1}{Academia Sinica Institute of Astronomy and Astrophysics, PO Box 23-141, Taipei 10617, Taiwan}
%\altaffiltext{2}{greg.schwarz@aas.org}

\begin{abstract}

We present measurements of the clustering properties of a sample of infrared (IR) bright dust-obscured galaxies (DOGs). 
Combining 125 deg$^2$ of wide and deep optical images obtained with the Hyper Suprime-Cam on the Subaru Telescope and all-sky mid-IR (MIR) images taken with {\it Wide-Field Infrared Survey Explorer}, we have discovered 4,367 IR-bright DOGs with $(i - [22])_{\rm AB}$ $>$ 7.0 and flux density at 22 $\micron$ $>$ 1.0 mJy.
We calculate the angular autocorrelation function (ACF) for a uniform subsample of 1411 DOGs with 3.0 mJy $<$ flux (22 $\micron$) $<$ 5.0 mJy and $i_{\rm AB}$ $<$ 24.0.
The ACF of our DOG subsample is well-fit with a single power-law, $\omega (\theta)$ = (0.010 $\pm$ 0.003) $\theta^{-0.9}$, where $\theta$ in degrees.
The correlation amplitude of IR-bright DOGs is larger than that of IR-faint DOGs, which reflects a flux-dependence of the DOG clustering, as suggested by \citet{Brodwin}.
We assume that the redshift distribution for our DOG sample is Gaussian, and consider 2 cases:
(1) the redshift distribution is the same as IR-faint DOGs with flux at 22 $\micron$ $<$ 1.0 mJy, mean and sigma $z$ = 1.99 $\pm$ 0.45, and (2) $z$ = 1.19 $\pm$ 0.30, as inferred from their photometric redshifts.
The inferred correlation length of IR-bright DOGs is $r_0$ = 12.0 $\pm$ 2.0 and 10.3 $\pm$ 1.7 $h^{-1}$ Mpc, respectively.
IR-bright DOGs reside in massive dark matter halos with a mass of $\log [\langle M_{\mathrm{h}} \rangle / (h^{-1} M_{\sun})]$ = 13.57$_{-0.55}^{+0.50}$ and 13.65$_{-0.52}^{+0.45}$ in the two cases, respectively.

\end{abstract}

\keywords{catalogs -- galaxies: active -- infrared: galaxies -- methods: statistical}



%======================
%     INTRODUCTION
%======================
\section{Introduction}
When and how did structures form and evolve in the 13.8-billion-year history of the Universe? 
This is one of the important questions we should solve in understanding the nature of galaxy formation and evolution. 
Investigating the clustering properties of galaxies as a function of redshift can give insights into the growth of structure and the relationship between galaxies and dark matter (DM).
The peak of star formation (SF) and active galactic nucleus (AGN) activity in the Universe and the bulk of stellar mass assembly in galaxies occurred at redshifts between 1 and 3, making it a particularly interesting epoch to study (e.g., \citealt{Richards,Goto,Bouwens,Madau,Guglielmo}).
One of the tools to investigate the clustering properties of photometrically selected samples of galaxies is the angular autocorrelation function (ACF).
Given the ACF and the galaxy redshift distribution in an adopted cosmology, and modeling the spatial correlation function as a power-law, we can determine the correlation length ($r_0$).
The clustering length in turn  depends on the bias of the galaxies relative to DM, allowing us to infer the mass of DM haloes in which they reside. 
Many studies of high-redshift ($z \sim$ 1 -- 3) populations such as quasars and submillimeter galaxies (SMGs) have found that they are strongly clustered ($r_0 \sim$ 10 $h^{-1}$ Mpc) as larger, and they reside in relatively massive DM halos ($M_{\rm h} \sim 10^{12-13} M_{\sun}$) (e.g., \citealt{Blain,Shen,Hickox_12,Allevato}, but see e.g., \citealt{Williams} for a cautionary remark for trying to measure the clustering of objects in small fields).

Here we focus on dust-obscured galaxies (DOGs: \citealt{Dey}) as a key population to understand the full picture of structure formation. 
DOGs are defined by their extreme optical and infrared (IR) color:
their mid-IR (MIR) flux densities are three orders of magnitude larger than those at optical wavelengths.
This extreme red color can be reproduced by active SF, AGN, or both, in which the bulk of the optical and ultraviolet (UV) emission is absorbed by surrounding dust which heats up and re-emits in the MIR.
It is well known that almost all massive galaxies in the present-day Universe harbor a supermassive black hole (SMBH) with a mass of $10^{6-10} M_{\sun}$ at their centers; their masses are strongly correlated with those of the spheroid component of their host galaxies.
This implies that galaxies and SMBHs coevolve (e.g., \citealt{Magorrian,Marconi}).
Galaxy merger models \citep{Hopkins_06} suggest that the central black hole and their host galaxies are obscured by a large amount of gas and dust during the initial stage of their co-evolution.
A hydrodynamic simulation conducted by \citet{Narayanan} found that when the BH accretion rate and SF rate (SFR) peak in a major merger, the object tends to be heavily dust-enshrouded, and thus appears as a DOG.
Therefore, investigating the clustering properties of DOGs could tell us how structures (i.e., DM halos) grow during the co-evolution of galaxies and SMBHs.

However, due to the optical faintness of DOGs, they have not been extensively studied in previous wide-field optical surveys. 
Thanks to the advent of small area ($<$ 10 deg$^2$) but deep optical surveys (e.g., the NOAO Deep Wide-Field Survey; \citealt{Jannuzi}) and deep IR data taken with {\it Spitzer}, over 3,000 DOGs have been discovered \citep{Dey}, and their physical and statistical properties have been investigated in detail (e.g., \citealt{Houck,Brand,Brodwin,Pope,Desai,Bussmann_09,Bussmann_11,Bussmann_12,Melbourne}). 
Such objects have typical flux density at 24 $\micron$ of $\sim$ 0.4 mJy; we refer to them in what follows as IR-``faint'' DOGs.
One of the important results of those studies relevant to this paper is that the redshift distribution of IR-faint DOGs is well-fit by a Gaussian, with mean and sigma $z$ = 1.99 $\pm$ 0.45 (e.g., \citealt{Dey,Desai,Yan}).

There is evidence that the clustering strength of DOGs depends on MIR flux, with the brighter IR DOGs being more strongly clustered.
\cite{Brodwin} suggested that IR brightest DOGs may be progenitors of previously unidentified present-day brightest cluster galaxies.
However, the clustering properties of DOGs with 22 $\micron$ flux greater than 1 mJy (hereinafter IR-``bright'' DOGs) are still unknown due to their low volume densities ($\Phi \sim 10^{-7}$ Mpc$^{-3}$: \citealt{Toba}) and thus small sample size in existing surveys.
Wide and deep optical imaging, together with wide and moderately deep IR imaging are needed to search for IR-bright DOGs.
The advent of capable imaging surveys at optical and IR wavelengths; Hyper Suprime-Cam (HSC: \citealt{Miyazaki}) on the Subaru Telescope and the {\it Wide-field Infrared Survey Explorer} ({\it WISE}: \citealt{Wright}), enables us to discover a large number of IR-bright DOGs. 
We hence performed a systematic search for IR-bright DOGs using the latest dataset of the HSC and {\it WISE}. 
The resulting large sample gives a statistically robust measurement of the clustering of IR-bright DOGs  for the first time.

This paper is organized as follows.
Section \ref{D_A} describes the sample selection of IR-bright DOGs, and how to derive the ACF.
The resultant ACF of IR-bright DOGs and comparison with that of IR-faint DOGs are presented in Section \ref{Results}.
 In Section \ref{Discussion}, we derive their correlation length, bias factor, and DM halo mass after discussing the possible uncertainties in the ACF mesurement, and compare them with other high-redshift populations. 
We also discuss their duty cycle by comparing the number density of DOGs with DM halo masses, and compare it with predictions from hydrodynamical simulations.
We summarize in Section \ref{Summary}.
The cosmology adopted in this paper assumes a flat Universe with $H_0$ = 68 km s$^{-1}$ Mpc$^{-1}$ ($h$ = 0.68), $\Omega_{\rm M}$ = 0.31, $\Omega_{\rm b}$ = 0.048, and $\Omega_{\Lambda}$ = 0.69, and the spectral index $n_{\rm s}$ = 0.96 and $\sigma_8$ = 0.83 \citep{Planck}. 
The correlation length is quoted in units of comoving $h^{-1}$ Mpc, and DM halo mass is quoted in units of $h^{-1} M_{\sun}$ with $H_0$ = 100 $h$ km s$^{-1}$ Mpc$^{-1}$ for ease of comparison with
previous works (All other physical quantities assume $h$ = 0.68).
All magnitudes refer to the AB system.

%======================
%   DATA AND ANALYSIS
%======================
\section{Data and analysis}
\label{D_A}
\begin{figure*}
 \begin{center}
 \includegraphics[width=0.8\textwidth]{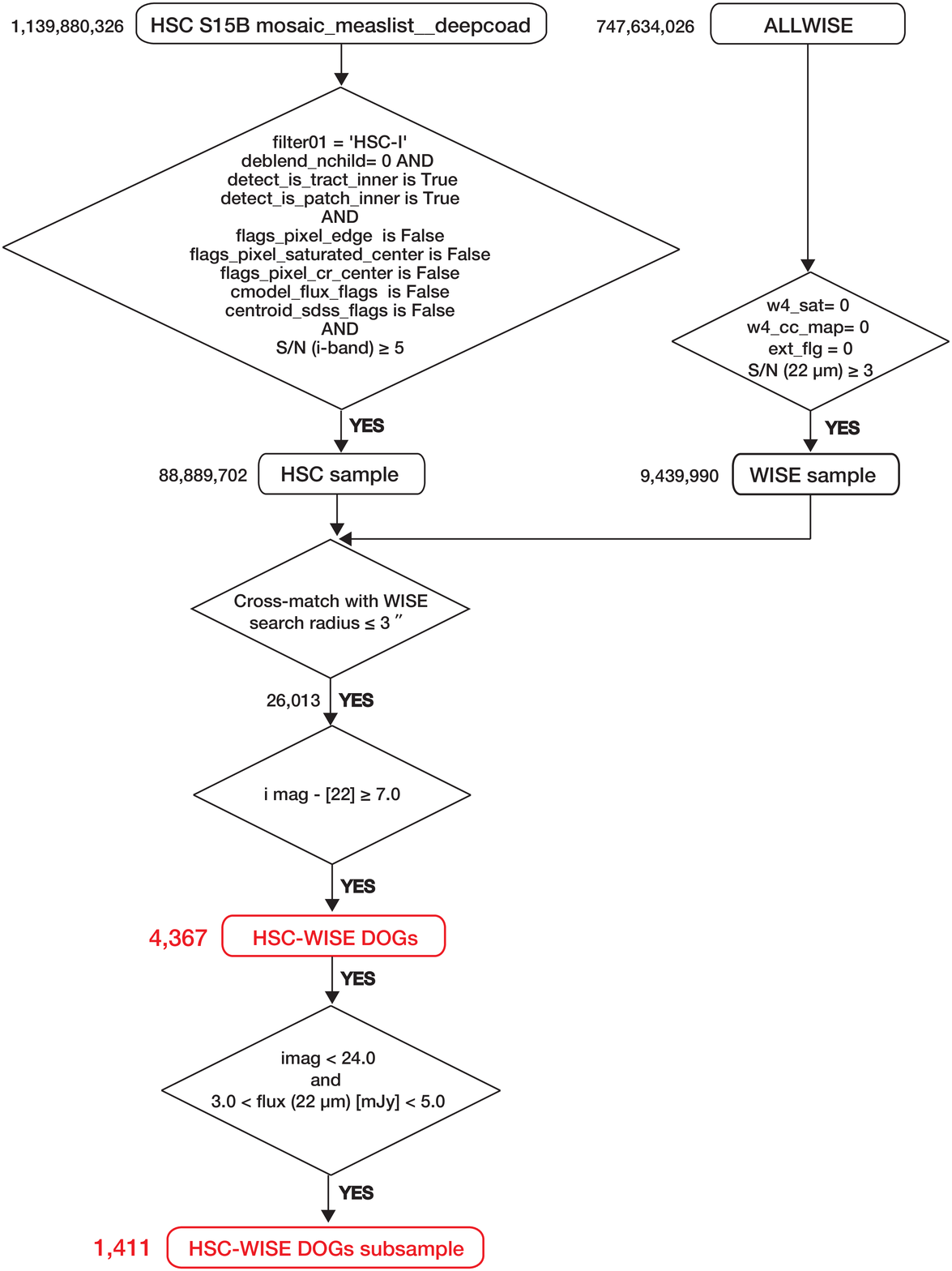}
 \end{center}
 \caption{Flow chart of the sample selection process.}
 \label{sample_selection}
\end{figure*}

We selected a DOG sample based on the {\it WISE} MIR catalog with optical counterparts detected by HSC.
Figure \ref{sample_selection} shows a flow chart of our sample selection process.
With this algorithm, we found a total of 4,367 IR-bright DOGs over 125 deg$^2$.\footnote{For the selection process, we employed the TOPCAT based on the Starlink Tables Infrastructure Library (STIL), which is an interactive graphical viewer and editor for tabular data \citep{Taylor}.}

%-----------------
%  Sample selection
%-----------------
\subsection{Sample selection}

%
%  HSC sample
%
\subsubsection{HSC sample}
\label{HSC}

The optical data were obtained with the Hyper Suprime-Cam (HSC: \citealt{Miyazaki}) on the 8.2 m Subaru telescope.
HSC is a CCD camera that covers a field of view 1.5 $\deg$ in diameter.  
The HSC Subaru Strategic Program (HSC-SSP\footnote{http://hsc.mtk.nao.ac.jp/ssp/}: Takada et al., in prep.) is now underway: we are carrying out a three-layered (wide, deep, and ultradeep) imaging survey  with five broad-band filters ($g$, $r$, $i$, $z$, and $y$).
This legacy survey started in March 2014, and 300 nights have been allocated to this project. 
It will cover a wide area ($\sim$ 1,400 deg$^2$) of the sky.

In this study, we utilized the HSC--SSP S15B data, containing the positions and photometric information of objects detected in observations from March 2014 to November 2015.
The total area of the survey footprint is about 125 deg$^2$.
We used a sample of 1,139,880,326 objects detected in the $g,r,i,z$ or $y$-band in the wide layer catalog.
The expected limiting magnitude (5$\sigma$, 2$\arcsec$ diameter aperture) and the median full width at half maximum (FWHM) of the point spread function (PSF) for $i$-band are $\sim$ 25.9 mag and $\sim$ 0.6 arcsec, respectively.
The data observed by the HSC in the S15B run were analyzed through the HSC pipeline (version 4.0.1; Bosch et al., in prep.) developed in conjunction with the Large Synoptic Survey Telescope (LSST) software pipeline \citep{Ivezic,Axelrod,Juric}.
This pipeline performs CCD-by-CCD reduction and calibration for astrometric and photometric zeropoints, mosaic-stacking which combines reduced CCD images into a single high signal-to-noise ratio (S/N) coadd image, and catalog generation for detecting and measuring sources on the coadd.
The photometric and astrometric calibrations are based on the data obtained from the Panoramic Survey Telescope and Rapid Response System (Pan-STARRS) 1 imaging survey \citep{Schlafly,Tonry,Magnier}.
In this work, we used the cModel magnitude to estimate $i$-band flux, which is a weighted combination of exponential and de Vaucouleurs fits to the light profile of each object (see \citealt{Lupton,Abazajian}).
Note that since our sample is located at relatively high galactic latitudes ($|b| >$ 20\degr), the influence of galactic extinction on the HSC photometry can be ignored.
We limited ourselves to those objects with clean $i$-band photometry with S/N $>$ 5 and removed duplicates in the same manner as \citet{Toba};
(i) objects are detected in $i$-band ({\tt filter01} = ``HSC-I''),
(ii) they are not blended or are children of blended objects ({\tt deblend\_nchild} = 0),
(iii) they are unique objects; repeat observations are removed ({\tt detect\_is\_tract\_inner} = ``True'' and {\tt detect\_is\_patch\_inner} = ``True''),
(iv) none of the pixels in their footprint are interpolated ({\tt flags\_pixel\_edge} = ``False''), 
(v) none of the central 3$\times$3 pixels are saturated ({\tt flags\_pixel\_saturated\_center} = ``False''),
(vi) none of the central 3$\times$3 pixels are affected by cosmic rays ({\tt flags\_pixel\_cr\_center} = ``False''),
(vii) there are no bad pixels in their footprint ({\tt flags\_pixel\_bad} = ``False''),
(viii) there are no problems in measuring cmodel fluxes ({\tt cmodel\_flux\_flags} = ``False''), and 
 (ix) they have a clean measurement of the centroid ({\tt centroid\_sdss\_flags} = ``False'').
Consequently, a sample of 88,889,702 sources is selected in 125 deg$^2$.
Note that the surface number density of the HSC sources is $\sim 7 \times 10^4$ deg$^{-2}$, which is larger than the earlier study of \cite{Toba}.
This is because improvements in the HSC pipeline which detects faint objects more reliably.

%
%  WISE sample
%
\subsubsection{WISE sample}
\label{WISE}

{\it WISE} has observed the whole sky at 3.4, 4.6, 12 and 22 $\micron$, with the PSF FWHM of 6.1, 6.4, 6.5, and 12.0 arcsec, respectively.
In this work, we utilized the latest ALLWISE catalog \citep{Cutri}.
The sensitivity at 3.4 and 4.6 $\micron$ in this catalog is better than the WISE all-sky data release \citep{Wright} because of improved data processing, while the number of 12 and 22 $\micron$--detected sources is smaller than in the earlier release because improved estimates of the local background reduced the number of faint objects.

We first limited ourselves to those objects with S\/N $>$ 3 at 22 $\micron$.
We then extracted point sources that are not affected by saturation, diffraction spikes, scattered-light halos, or optical ghosts by taking the saturation flag, extend flag, and image artifact flag into account,
leaving a sample with clean photometry.
In this work, we used the profile-fit magnitude for each band, which is optimized for point sources and thus provides reliable photometry for our sample.
We converted the {\it WISE} Vega system magnitudes into AB magnitude, by adding 2.699, 3.339, 5.174, and 6.620 to the Vega magnitude at 3.4, 4.6, 12, and 22 $\micron$, respectively, according to the Explanatory Supplement to the AllWISE Data Release Products \footnote{http://wise2.ipac.caltech.edu/docs/release/allwise/expsup/}.
Consequently, a sample of 9,439,990 sources is selected over the whole sky.

%
% Cross-identification of HSC and WISE data
%
\subsubsection{Cross-identification of HSC and WISE data}
\label{Cross}
\begin{figure}
 \begin{center}
 \includegraphics[width=0.45\textwidth]{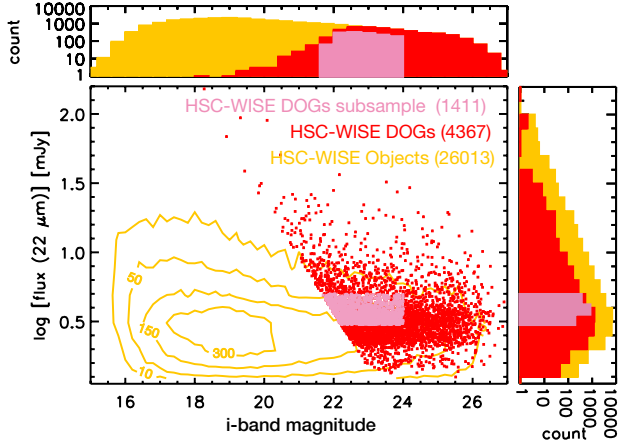}
 \end{center}
 \caption{The distribution of $i$-band magnitudes and 22 $\micron$ fluxes of our sample. The histograms of $i$-band magnitude and 22 $\micron$ flux are given on the top and right, respectively. The yellow contours represent the number density of the HSC--{\it WISE} objects (26,013 objects in total) in each 0.2 $\times$ 0.2 region on the $i$-band magnitude--log [flux (22 $\micron$)] plane. The points and histograms with red and pink color represent the 4,367 HSC-{\it WISE} DOGs and 1,411 HSC-{\it WISE} DOGs subsamples, respectively.}
 \label{mag_flux}
\end{figure}

We cross-identified the HSC sample with the {\it WISE} sample using a matching radius of 3 arcsec, which yields 26,013 matches (hereinafter HSC--{\it WISE} objects).
We then applied the DOGs selection of $i - [22] >$ 7.0, where $i$ and [22] represent AB magnitudes in the HSC $i$-band and {\it WISE} 22 $\micron$ band, respectively \citep{Toba}.
Note that this threshold is consistent with the original DOG definition, $R - [24] > 14$ in Vega magnitudes (i.e., $R - [24 ] > 7.5$ in AB magnitudes), as we found based on the DOG sample in \cite{Melbourne} (see \citealt{Toba}).
This yields a sample of 4,367 IR-bright DOGs (hereinafter HSC-{\it WISE} DOGs), the largest sample of such objects in the literature.
Figure \ref{mag_flux} shows the 22 $\micron$ flux and $i$-band magnitude distributions for our sample. 
The average and median 22 $\micron$ flux densities of all HSC-{\it WISE} matched objects (HSC-{\it WISE} objects) are 4.21 and 3.39 mJy, respectively, while their average and median $i$-band magnitudes are 19.8 and 19.4, respectively.
The average and median 22 $\micron$ flux densities of HSC-{\it WISE} DOGs are 4.00 and 3.27 mJy, respectively, while their average and median $i$-band magnitudes are 23.5 and 23.3, respectively.
About 27.1\% of {\it WISE} sources have more than one HSC counterpart when selecting HSC-{\it WISE} objects since the spatial resolution of the HSC is much better than that of {\it WISE}. 
In this study, we chose the closest object as the optical counterpart.
However, this does not necessarily mean that the nearest source is always the real counterpart.
We will discuss how the nearest matching influences the clustering properties of IR-bright DOGs in Section \ref{effect_nearst}.

We also note that our DOG selection procedure differs from that of \cite{Toba} who discovered IR-bright DOGs using HSC early data (S14A\_0) and ALLWISE data.
To reduce the uncertainties of the identification of DOGs given the poor resolution of {\it WISE}, \cite{Toba} first joined the HSC data with near-IR (NIR) data obtained from the VISTA Kilo-degree Infrared Galaxy survey (VIKING: \citealt{Arnaboldi}) whose angular resolution is roughly comparable to that of the HSC.
They then adopted an optical -- NIR color cut to the merged catalog to reject sources unlikely to be DOGs before cross-matching with {\it WISE}.
They thus significantly reduced the fraction of {\it WISE} sources with multiple HSC counterparts, although about 7\% of {\it WISE} sources still had more than one HSC counterpart.
On the other hand, this method could miss NIR-faint DOGs that are not detectable by VIKING, giving rise to bias when comparing the clustering properties of our IR-bright DOG sample with those of IR-faint DOGs. 
We compared the number count of DOGs selected in \cite{Toba} and this work.
We found that the shape of the number counts as a function of magnitude are similar in the two cases but the number of HSC-{\it WISE} DOG sample in this work is about 10-20 \% larger than that in \citet{Toba}, which could mean that some DOGs in this work are mis-identified.

In order to measure the clustering of our sample, we then created a flux-limited subsample, with uniform depth throughout the survey area.
We narrowed our DOG sample to 1,411 objects with $i$-band magnitude $<$ 24.0 and 3.0 mJy $<$ $F$ (22 $\micron$) $<$ 5.0 mJy.
Note that the upper threshold for flux at 22 $\micron$ removes extremely IR-bright DOGs with 22 $\micron$ flux $>$ 5.0 mJy, that have a significantly different redshift distribution from the fainter objects \citep{Toba_16}.
While we have removed optically fainter DOGs, our DOG sample objects are still considerably fainter than the SDSS limiting magnitude.
The combination of deep and wide survey data with HSC and {\it WISE} provides us a spatially rare IR-bright DOG sample that has been undetected by previous surveys.
Hereinafter, we investigate the clustering properties for this subsample of 1,411 HSC-{\it WISE} DOGs.

%-----------------
% The two-point autocorrelation function of IR-bright DOGs
%-----------------
\subsection{The angular autocorrelation function of IR-bright DOGs}
We derive the angular autocorrelation function (ACF) for the 1,411 IR-bright DOGs discovered in this work. 
We adopt the Landy-Szalay estimator \citep{Landy} which can be formulated as
\begin{equation}
\omega_{\rm obs} (\theta) = \frac{DD (\theta) - 2 DR (\theta) + RR (\theta)}{RR (\theta)},
\label{LS}
\end{equation}
where $DD$, $DR$, and $RR$ are the ``normalized'' numbers of data-data, data-random, and random-random pairs in each separation bin ($\theta$) in degree (hereinafter, $\theta$ is given in degrees), respectively.
Those are defined as follows:
\begin{eqnarray}
DD (\theta) &=& n_{{\rm DD}} (\theta) \left[ \frac{N_{\rm D} (N_{\rm D} -1)}{2} \right]^{-1},\\
DR (\theta) &=&  \frac{n_{\rm DR (\theta)}}{N_{\rm D}N_{\rm R}},  \\
RR (\theta) &=& n_{{\rm RR}} (\theta) \left[ \frac{N_{\rm R} (N_{\rm R} -1)}{2} \right]^{-1},
\end{eqnarray}
where $n_{\rm DD}$, $n_{\rm DR}$, and $n_{\rm RR}$ are the actual numbers of data-data, data-random, and random-random pairs in each separation bin ($\theta$), respectively.
$N_{\rm D}$ and $N_{\rm R}$ are the total number of DOGs and random data, respectively.
The uncertainty of the ACF ($\sigma_{\omega_{\rm obs}}(\theta)$) is calculated by the bootstrap method as
\begin{equation}
\sigma_{\omega_{\rm obs}} (\theta) = \sqrt{\frac{1}{N-1}\sum_{i=1}^N \,[\omega_i (\theta) - \bar{\omega} (\theta)]^2},
\end{equation}
where $N$ is the number of bootstrap resamples and $\bar{\omega} (\theta)$ is defined as
\begin{equation}
\bar{\omega} (\theta) = \frac{1}{N}\sum_{i=1}^N \omega_i (\theta)
\end{equation}
\citep{Ling}.
In this work, we used $N$ = 1000.
In practice, we create a random sample composed of 100,000 sources with the same geometrical constraints as the data sample. 
The random sample watches the HSC geometry, is produced by a module \footnote{https://github.com/jcoupon/maskUtils} developed for the HSC, which we take into account the flags described in Section \ref{HSC}.
We then distributed these points on the {\it WISE} images and removed those that overlap image artifacts such as diffraction spikes, scattered-light halos, and optical ghosts, following the {\it WISE} sample selection as described in Section \ref{WISE}.
The positions affected by artifacts are described in the Explanatory Supplement to the AllWISE Data Release Products.

Note that we cannot measure the contribution from fluctuations on larger scales than our survey area.
Hence, the real correlation function is offset from the observed function by an integral constraint (IC; \citealt{Groth});
\begin{equation}
\omega (\theta) = \omega_{\rm obs} (\theta)  + {\rm IC},
\end{equation}
where IC is given by
\begin{equation}
{\rm IC} = \frac{\sum [RR (\theta) A_{\omega}\theta^{-0.9}]}{\sum RR (\theta)}
\label{IC}
\end{equation}
\citep{Roche}.
In this work, given the value of $A_{\omega}$ derived below, we obtained IC = 0.008.
The ACF can be approximated by a power-law form as follows:
\begin{equation}
\label{Aw_real}
\omega (\theta) = A_{\omega} \theta^{-\beta}.
\end{equation}
where $A_{\omega}$ and $\beta$ are the correlation amplitude and the power-law index.
Since one of the purposes of this study is to compare the correlation amplitude with IR-faint DOGs, we followed \cite{Brodwin} and fixed $\beta$ to be 0.9.

%======================
%     RESULTS
%======================
\section{Results}
\label{Results}

Here we present the ACF and compare the correlation amplitude of IR-bright DOGs with those of IR-faint DOGs.
\begin{figure}
 \begin{center}
 \includegraphics[width=0.45\textwidth]{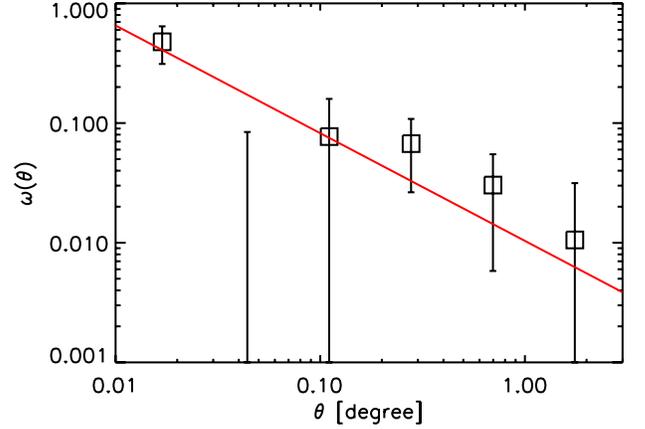}
 \end{center}
 \caption{Two-point angular correlation function of IR-bright DOGs. The red line represents the best fit power-law with Equation (\ref{Aw_real}).}
 \label{ACF}
\end{figure}
Figure \ref{ACF} shows the resultant ACF using Equation (\ref{LS})--(\ref{IC}), for IR-bright DOGs.
A fit to Equation (\ref{Aw_real}) with $\beta = 0.9$ gives $A_{\omega}$ = 0.010 $\pm$ 0.003.
\begin{figure*}
 \begin{center}
 \includegraphics[width=0.8\textwidth]{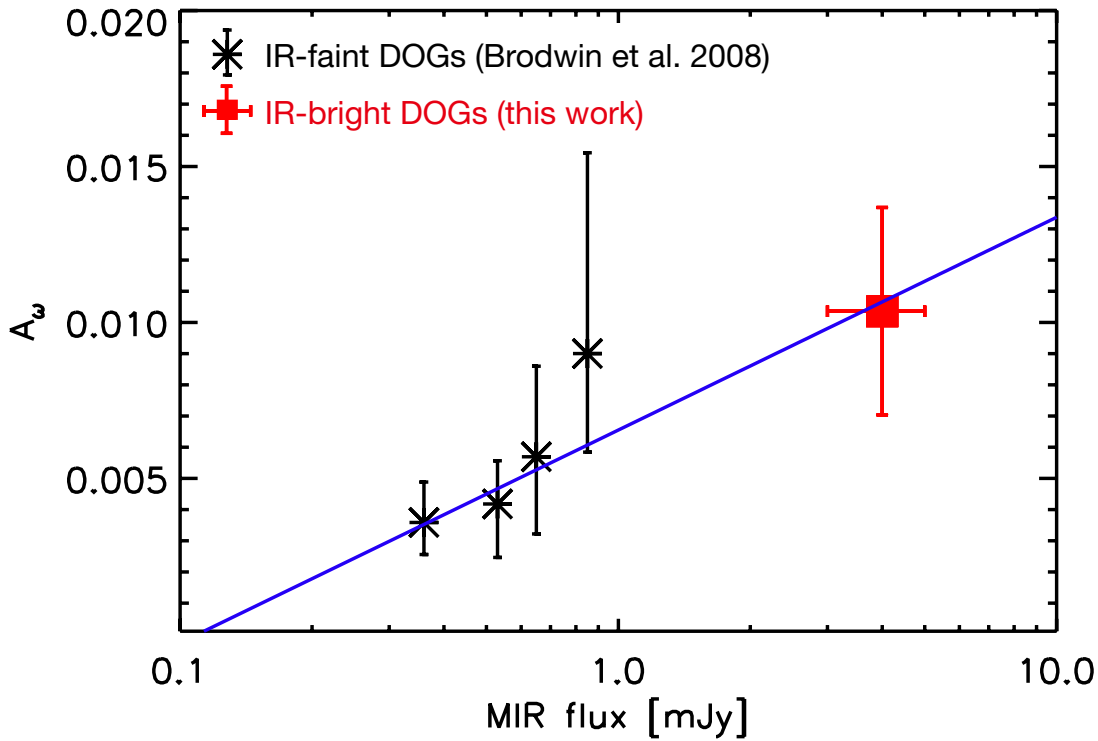}
 \end{center}
\caption{Correlation amplitude ($A_{\omega}$) as a function of MIR flux. Asterisks represent the correlation amplitude of IR-faint DOGs calculated from the correlation length presented by \citet{Brodwin},  while the red square represents that of IR-bright DOGs (this work). The blue line shows the best-fit line with $A_{\omega} = 0.0072 \log (f_{\rm MIR}) + 0.0067$ (Equation (\ref{Aw_fit})). } 
\label{Aw}
\end{figure*}
Figure \ref{Aw} compares this value with that of IR-faint DOGs where we estimated $A_{\omega}$ for IR-faint DOGs based on the correlation length ($r_0$) derived from \citet{Brodwin}.
Note that they estimated the correlation length as a function of {\it Spitzer} 24 $\micron$ flux, but the difference from the {\it WISE} 22 $\micron$ flux we use is negligible.
We found that (i) the correlation amplitude of IR-bright DOGs is consistent with that at the bright end ($F_{\rm MIR}$ $\sim$ 1.0 mJy) of IR-faint DOGs and (ii) the correlation amplitude increases with increasing MIR flux.
We fit the data as 
\begin{equation}
\label{Aw_fit}
A_{\omega} =  (7.2^{+1.9}_{-2.2})\times 10^{-3} \, \log (f_{\rm MIR}) + (6.7^{+0.8}_{-0.7})\times 10^{-3},
\end{equation}
where $f_{\rm MIR}$ is given in mJy.
Our results use DOGs to $F$ (22 $\micron$) $=$ 5.0 mJy, and thus Equation (\ref{Aw_fit}) is applicable only to that flux.

%======================
%    DISCUSSIONS
%======================
\section{Discussion}
\label{Discussion}

In this section, we discuss the clustering properties of our subsample of 1,411 HSC-{\it WISE} DOGs.
Assuming a model for their redshift distribution, we derive their correlation length, bias factor, and dark matter halo mass.
However, the redshift distribution of our sample has not yet been constrained spectroscopically.
We thus consider two possible models for their redshift distribution and derive their clustering properties for each case.
We then discuss the duty cycle of IR-bright DOGs and compare it with theoretical predictions.

%-----------------
% Influence of the nearest matching on the correlation length
%-----------------
\subsection{Influence of the optical--IR matching algorithm on the correlation amplitude}
\label{effect_nearst}
In Section \ref{Cross}, we simply selected DOG candidates by finding the nearest HSC source for each {\it WISE} source. 
But the nearest source may not always be the true counterpart.
Indeed, about 27\% of {\it WISE} sources have more than one HSC counterpart within 3\arcsec, which is far from negligible. 
We hence investigate the influence of the nearest matching on derived correlation amplitude.
Here we estimate this effect using Monte Carlo simulations. 
When a {\it WISE} source has multiple counterpart candidates within the 3\arcsec \,search radius, we randomly choose one object and determine if it satisfies our selection criteria.
We then calculate the ACF and correlation amplitude for the resulting DDG sample.
We create 1000 such realizations and estimate the distribution of $A_{\omega}$. 
The mean value and standard deviation of the simulated correlation amplitude are $A_{\omega}$ = 0.020 $\pm$ 0.010, which is roughly consistent with our result above.
Therefore, we conclude that the influence of the uncertainty due to our nearest matching algorithm on the resultant correlation amplitude is modest.

%-----------------
% Redshift distribution of IR-bright DOGs
%-----------------
\subsection{Redshift distribution of IR-bright DOGs}
\label{effect_z}

\begin{figure}
 \begin{center}
 \includegraphics[width=0.45\textwidth]{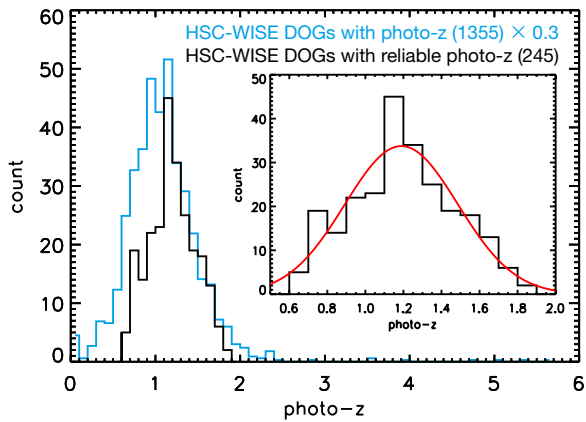}
 \end{center}
 \caption{Photometric redshift distribution of IR-bright DOGs inferred from {\tt MIZUKI} \citep{Tanaka}. The cyan line represents the photometric redshift distribution of 1,355 DOGs without the quality cuts (scaled down by a factor of $\sim$ 3). The black line and the inserted figure show the photometric redshift distribution of 245 DOGs sample with reduced $\chi^2$ $<$ 1.5 and relative error of photometric redshift is less than 5 \%. The distribution can be fitted with a Gaussian, with mean and sigma $z$ = 1.19 $\pm$ 0.30 as shown as the red solid line.}
 \label{photo-z}
\end{figure}

To derive the correlation length, bias factor, and dark matter halo of the subsample of HSC-{\it WISE} DOGs, we need to know their redshift distribution, $N (z)$.
Since the redshift distribution of our sample has not yet been constrained spectroscopically, we here consider two possible cases.\\
{\bf Case 1}: Following \citet{Brodwin}, we assume that $N (z)$ is Gaussian, with mean and sigma $z$ = 1.99 $\pm$ 0.45, as was determined from the IR-faint DOG sample in \cite{Dey}.
That sample was selected by adopting $R - [24] > 14$ and $F\,(24 \, \micron) > 0.3$ mJy, where $R$ and [24] represent Vega magnitudes in the $R$-band and {\it Spitzer} 24 $\micron$, respectively.
\cite{Dey} measured spectroscopic redshifts for 86 IR-faint DOGs, although the objects with redshifts  
were largely from the bright end of their sample.
Since clustering properties such as correlation length are sensitive to the given $N (z)$, this assumption is useful when comparing the clustering properties of IR-bright DOGs with those of IR-faint DOGs under the same condition.\\
{\bf Case 2}: We assume that $N (z)$ is Gaussian, with mean and sigma $z$ = 1.19 $\pm$ 0.30 determined by the photometric redshifts of the galaxies in our sample.
Our DOG sample is too faint in the optical to have spectra in SDSS \citep{Alam} or the Galaxy And Mass Assembly survey (GAMA: \citealt{Driver_09,Driver_11}).
We thus attempted to infer photometric redshifts of IR-bright DOGs with a custom-designed Bayesian photometric redshift code ({\tt MIZUKI}: \citealt{Tanaka}) using the 5-band HSC photometry.
The photometric redshifts are computed for objects with clean cModel photometry based on the SED fitting technique where the spectral templates of galaxies are generated with the \cite{Bruzual} code.
1355/1411 ($\sim$ 96.0\%) objects have photometric redshifts, and their distribution is shown in Figure \ref{photo-z}.
But only  245 / 1411  (17.3 \%) of those redshifts are reliable in the sense that the reduced $\chi^2_{\rm z} <$ 1.5 and $\sigma_{\rm z} / z < $ 5\%.
Note that we are using $z$ here for redshift.
We confirmed that these 245 objects are distributed uniformly in the $i$-band magnitude and 22 $\micron$ flux plane (see Figure \ref{mag_flux}), meaning that the estimate of their photometric redshift distribution is not affected by possible flux dependence of the photometric redshift.

\begin{figure*}
 \begin{center}
 \includegraphics[width=\textwidth]{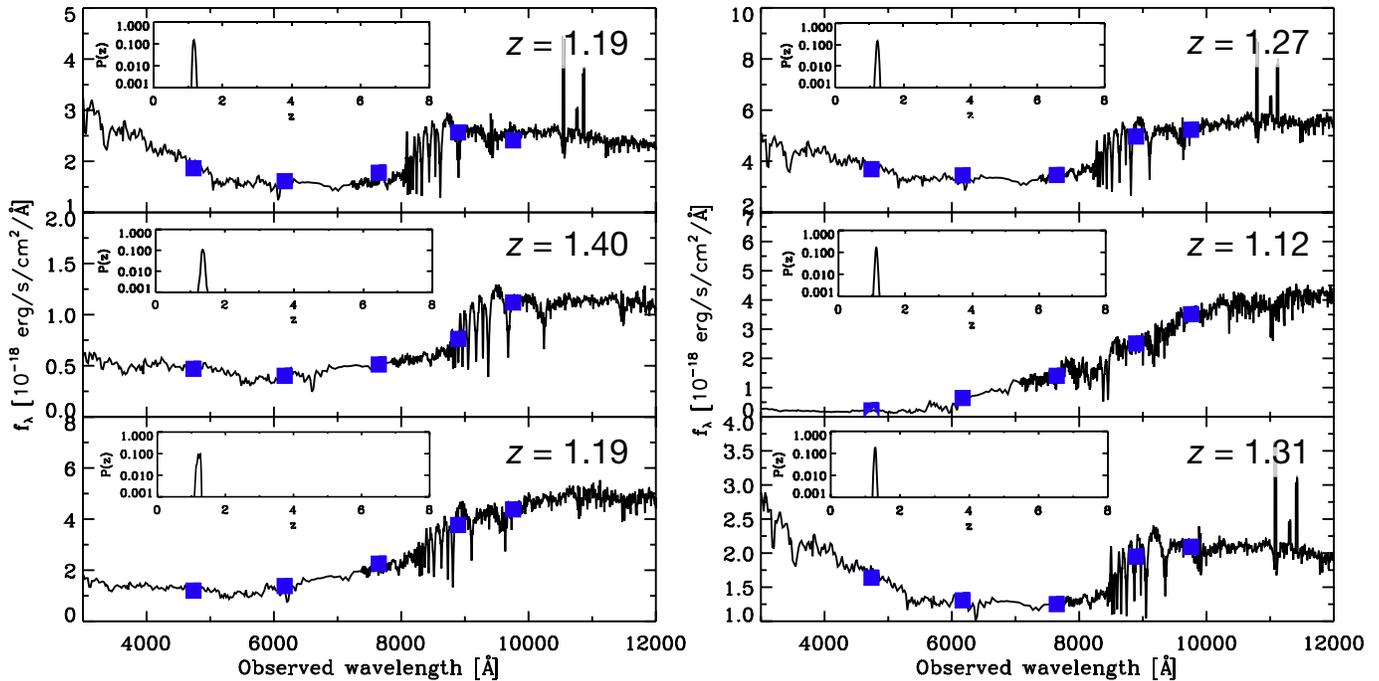}
 \caption{Example of SED fitting for 6 DOGs. Blue squares represent the data points while the black lines represent the best-fit SED templates \citep{Tanaka}. The inserted panel shows the probability distribution of redshift, $P(z)$.}
 \label{SED}		  
 \end{center}
\end{figure*}

Their photometric redshift distribution is well-fit by a Gaussian, with mean and sigma $z$ = 1.19 $\pm$ 0.30, which is lower than the redshift distribution assumed in Case 1.
Note that the sigma of the redshift distribution is much larger than the photometric redshift errors.
Figure \ref{SED} shows examples of SED fits for 245 DOGs based on {\tt MIZUKI}.
We found that the photometric redshifts of objects in the sample are determined from the Balmer break feature at 4000\AA .
Also, the probability distribution function of photo-z ($P(z)$) is narrow and has no secondary peak in every case, which ensures the reliability of our measurement of photo-z.
At the same time, however, the redshift distribution could be biased to $z < 1.5$ because one cannot determine reliable photometric redshifts for objects whose Balmer break lies beyond the $y$-band (e.g., star forming galaxies at $z >$ 1.5) or with power-law SEDs (e.g., dusty AGNs) and thus these objects may not be included in 245 HSC-{\it WISE} DOGs photo-z sample.
Nevertheless, in what follows, we calculate correlation length and related quantities assuming both Case 1 and Case 2 redshift distributions, and tabulate them in Table \ref{Table}.

\floattable
\begin{deluxetable}{ccccccccccc}
\tablecaption{Clustering properties of IR-bright DOGs.\label{Table}}
\tablecolumns{11}
\tablenum{2}
\tablewidth{0pt}
\tablehead{
\colhead{Case} & \colhead{flux (22 $\micron$)  } & \colhead{$N$} & \colhead{$z$ interval} & 
\colhead{$\bar{z}$} & \colhead{$A_{\omega}$} & \colhead{$r_0$} & $b$ & \colhead{$\log (M_{\rm h})$} &
\colhead{$\log (M_{\rm h, min})$} & \colhead{$f_{\rm duty}$}\\
\colhead{} & \colhead{[$mJy$]} & \colhead{} & \colhead{} & \colhead{}  & \colhead{} & \colhead{$h^{-1}$ Mpc}  & \colhead{} & \colhead{$h^{-1}$ $M_{\sun}$} & \colhead{$h^{-1}$ $M_{\sun}$}
}
\startdata
1 & 3.0 $< f_{22} <$ 5.0 & 1,411 & 1.54 $< z <$ 2.44 & 1.99 & 0.010 $\pm$ 0.003 & 12.0 $\pm$ 2.0 & 5.99 $\pm$ 0.96 & 13.57$_{-0.55}^{+0.50}$ & 13.40$_{-0.59}^{+0.53}$ & $>$ 0.013\\	  
2 &  3.0 $< f_{22} <$ 5.0 & 1,411 & 0.89 $< z <$ 1.49 & 1.19 & 0.010 $\pm$ 0.003 & 10.3 $\pm$ 1.7 & 3.88 $\pm$ 0.62 & 13.65$_{-0.52}^{+0.45}$ & 13.41$_{-0.56}^{+0.49}$ & $>$ 0.007\\	
\enddata
%\tablenotetext{a}{At exposure start.}
\tablecomments{Assuming the two cases of the redshift distribution (case 1 and 2), we estimated the correlation length ($r_0$), bias factor ($b$), and the dark matter halo masses for each case (see the main text for details).}
\end{deluxetable}

%-----------------
% Correlation Length
%-----------------
\subsection{Correlation length}
\label{corre}
\begin{figure*}
 \begin{center}
 \includegraphics[width=0.8\textwidth]{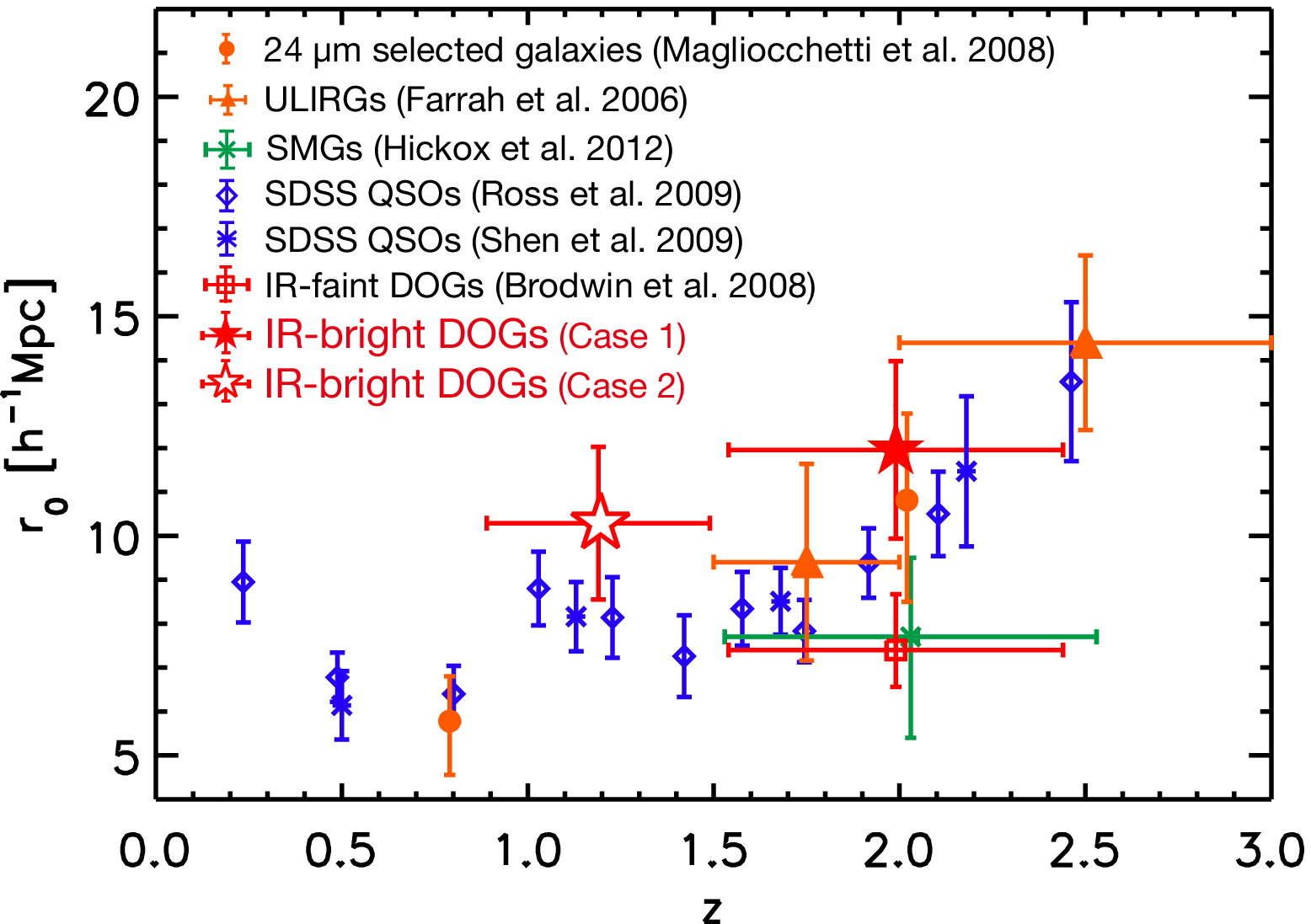}
 \end{center}
 \caption{Correlation length as a function of redshift. Blue diamonds and asterisks represent the correlation lengths of quasars selected from the SDSS DR5 \citep{Ross,Shen}. The green cross represents that of SMGs \citep{Hickox_12}. The orange triangles represents that of ULIRGs \citep{Farrah} while the orange circles represent that of 24 $\micron$ selected galaxies \citep{Magliocchetti}. The open red square represents that of IR-faint DOGs \citep{Brodwin} while red filled and open stars represent the correlation length of IR-bright DOGs, assuming the $N(z)$ for Case 1 and 2, respectively.}
\label{r0}
\end{figure*}

The ACF determined by Equation (\ref{Aw_real}) can be deprojected \citep{Limber} to yield a measurement of the real-space correlation length ($r_0$) over the redshift range;
\begin{equation}
r_0 = \left[ A_{\omega} \frac{c}{H_0 H_{\gamma}}\frac{[\int^{z_{\rm max}}_{z_{\rm min}} N(z) {\rm d}z]^2}{\int^{z_{\rm max}}_{z_{\rm min}} N(z)^2 \chi(z)^{1-\gamma} E(z) {\rm d}z} \right]^{\frac{1}{\gamma}},
\label{Lim}
\end{equation}
where $\gamma$, $H_{\gamma}$, $E(z)$, and $\chi(z)$ are defined as follows:

\begin{equation}
\label{ganma}
\gamma = 1 + \beta,
\end{equation}

\begin{equation}
H_{\gamma} = \frac{\Gamma (\frac{1}{2}) \Gamma(\frac{\gamma -1 }{2})}{\Gamma (\frac{\gamma}{2})},
\end{equation}

\begin{equation}
E(z) = \sqrt{\Omega_M (1+z)^3 + \Omega_{\Lambda}},
\end{equation}
and
\begin{equation}
\chi (z) = \frac{c}{H_0} \int^z_0 \frac{1}{E(z^{\prime})} {\rm d}z^{\prime}.
\label{chi0}
\end{equation}
$N(z)$ is the redshift distribution of DOGs.
Assuming that the redshift distribution for our DOG sample is Gaussian, with mean and sigma $z$ = 1.99 $\pm$ 0.45 (Case 1) and $z$ = 1.19 $\pm$ 0.30 (Case 2), we derived the correlation length ($r_0$) from Equation (\ref{Lim}) -- (\ref{chi0}).
We found $r_0$ = 12.0 $\pm$ 2.0 and 10.3 $\pm$ 1.7 $h^{-1}$ Mpc for Case 1 and 2, respectively.

Figure \ref{r0} represents the comparison of the correlation length of our DOG sample with other high-redshift strongly-clustered populations; SDSS quasars \citep{Ross,Shen}, IR-selected galaxies/AGNs/ultraluminous IR galaxies (ULIRGs) \citep{Farrah,Magliocchetti}, SMGs \citep{Hickox_12}, and IR-faint DOGs \citep{Brodwin}.
These different papers use somewhat different cosmological models; we estimate that the resulting systematic uncertainty on the correlation length is $\sim$ 3\%.

Figure \ref{r0} indicates that the correlation length of IR-bright DOGs is comparable to or larger than those of other populations of comparable redshift, for either of our assumptions about $N(z)$ 
(although the correlation length error bars overlap those of ULIRGs and SDSS quasars).
In addition, the correlation length of IR-faint DOGs differs significantly from that of IR-bright DOGs.
Although \citet{Brodwin} reported that the correlation length of IR-faint DOGs is consistent with that of SMGs and quasars, the correlation length of our IR-bright DOG sample is similar to or even larger than quasars at same redshift.
This indicates that the IR-bright DOGs are a different population from IR-faint DOGs and they do not have an evolutionary link at least with SMGs at similar redshift (see Section \ref{BF}).

%-----------------
% Bias factor of IR-bright DOGs
%-----------------
\subsection{Bias factor of IR-bright DOGs}
\label{BF}
Here we estimate the bias factors assuming the two redshift distributions; $z$ = 1.99 $\pm$ 0.45 (Case 1) and 1.19 $\pm$ 0.30 (Case 2).
The spatial correlation function of galaxies is usually expressed by a power law as
\begin{equation}
\xi (r) = \left( \frac{r}{r_0} \right)^{-\gamma},
\end{equation}
where $\gamma$ = 1.9.
From the real-space correlation function, we define the DOGs dark matter clustering bias ($b_{\rm DOG}$) at  large scales (8 $h^{-1}$ Mpc) as 
\begin{equation}
\label{bias}
b_{\rm DOG} = \sqrt{\frac{\xi_{\rm DOG} (8,z)}{\xi_{\rm DM} (8,z)}},
\end{equation}
where $\xi_{\rm DOG} (8,z)$ and $\xi_{\rm DM} (8,z)$ are the two-point correlation function of DOGs and of the underlying dark matter, respectively.
We estimated them in the same manner as \citet{Ikeda} (see also \citealt{Peebles,Myers});
\begin{equation}
\xi_{\rm DOG} (8,z) = \left( \frac{r_0}{8} \right)^{\gamma}
\end{equation}
and
\begin{equation}
\xi_{\rm DM} (8,z) = \frac{(3-\gamma)(4-\gamma)(6-\gamma) 2^{\gamma}}{72} \left[\sigma_8 \frac{g(z)}{g(0)}\frac{1}{1+z} \right]^2,
\end{equation}
where

\begin{equation}
g (z) = \frac{5 \Omega_{\rm mz}}{2} \left \{ \Omega_{\rm mz}^{\frac{4}{7}} - \Omega_{\rm \Lambda z} +\left(1+\frac{\Omega_{\rm mz}}{2} \right)  \left(1+\frac{\Omega_{\rm \Lambda z}}{70} \right)  \right \}^{-1}
\end{equation}
and
\begin{eqnarray}
\Omega_{\rm mz} &=& \frac{\Omega_M (1+z)^3}{\Omega_M (1+z)^3 + \Omega_{\Lambda}} \\
\Omega_{\Lambda z} &=& \frac{\Omega_{\Lambda}}{\Omega_M (1+z)^3 + \Omega_{\Lambda}}.
\label{Omega_l}
\end{eqnarray}
Using Equation (\ref{bias}) - (\ref{Omega_l}), we find the bias factors of IR-bright DOGs to be $b_{\rm DOG}$ = {5.99 $\pm$ 0.96 and 3.88 $\pm$ 0.62} for Case 1 and 2, respectively. 

\begin{figure*}
 \begin{center}
 \includegraphics[width=\textwidth]{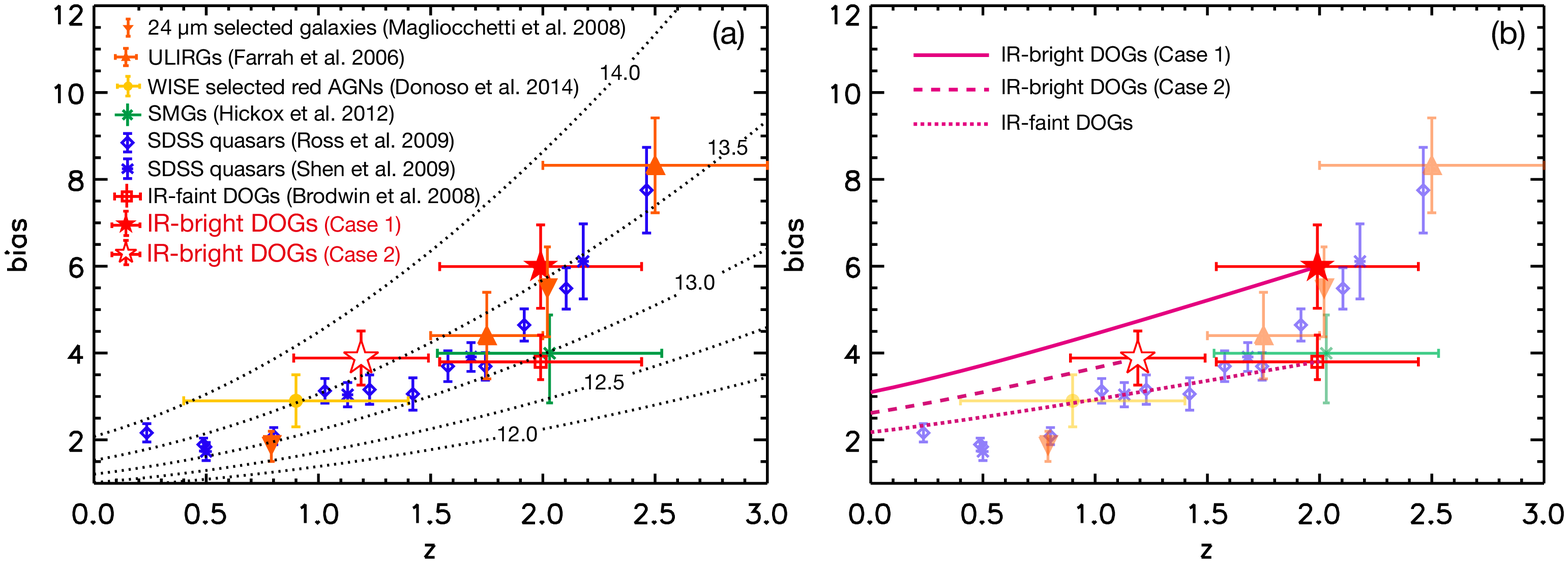}
 \end{center}
 \caption{(a) Bias factors as a function of redshift. Blue diamonds and asterisks represent the bias factors of quasars selected from the SDSS DR5 \citep{Ross,Shen}. The green cross represents that of SMGs \citep{Hickox_12}. The orange triangles represent that of ULIRGs \citep{Farrah} while the orange upside down triangles represent that of 24 $\micron$ selected galaxies \citep{Magliocchetti}. The yellow circle represents the bias factor of {\it WISE}-selected red AGN with  ($r$ - [3.4])$_{\rm Vega}$ $>$ 6 \citep{Donoso}. The red open square represents that of IR-faint DOGs \citep{Brodwin} while red stars represent the bias factor for IR-bright DOGs in this work. The dashed lines show the evolution of halo bias factor  for fixed DM halo mass of $\log [\langle M_{\mathrm{h}} \rangle / (h^{-1} M_{\sun})]$ = 12.5, 13.0, 13.5 and 14.0 from bottom to top, respectively using Equation (\ref{d_M}). (b) Symbols are same as (a). The lines show the evolution of bias for a passive population of tracers \citep{Fry,White}.}
\label{bias_comp}
\end{figure*}

In Figure \ref{bias_comp} we plot the estimated bias factors with those of other populations; SDSS quasars \citep{Ross,Shen}, IR-selected galaxies/AGNs/ULIRGs \citep{Farrah,Magliocchetti,Donoso}, SMGs \citep{Hickox_12}, and IR-faint DOGs \citep{Brodwin}, as a function of redshift.
We have re-calculated the bias factors using the same cosmology from their correlation lengths if available in the literature.
Note that the re-calculated bias factors of the SDSS quasars are larger that those in original papers, because of the difference of the assumed cosmology and method of the estimation.
In case 1, the bias factor of IR-bright DOGs is consistent with that of ULIRGs and perhaps SDSS quasars, while the bias factor of IR-bright DOGs is consistent with that of SDSS quasars, as shown in Figure \ref{bias_comp}.
Since most DOGs satisfy the definition of ULIRGs ($L_{\rm IR} > 10^{12}\, L_{\sun}$) (e.g., \citealt{Dey,Melbourne,Toba}), the fact that IR-bright DOGs and ULIRGs have similar bias factor is reasonable.

%-----------------
% Dark matter halo mass of IR-bright DOGs
%-----------------
\subsection{Dark matter halo mass of IR-bright DOGs} 

Using an ellipsoidal collapse model, \citet{Sheth} related the halo bias factor to its mass and calibrated a fitting relation for a large library of cosmological N-body simulations;
\begin{equation}
{\scriptstyle 
b (M,z) = 1 + \frac{1}{\sqrt{a} \delta_c} \left[ a \nu^2 \sqrt{a} + b\sqrt{a}(a\nu^2)^{1-c} - \frac{(a\nu)^2}{(a\nu^2)^c + b(1-c)(1-\frac{c}{2})} \right],
\label{d_M}
}
\end{equation}
where a = 0.707, b = 0.5, c = 0.6,
$\delta_c$ = 1.686 is the critical overdensity required for collapse and $\nu = \frac{\delta_c}{\sigma (M) D(z)}$.
The variance of mass fluctuations inside a sphere of radius $R$ at $z=0$, $\sigma^2 (M)$, is given by
\begin{equation}
\label{sigma_M}
\sigma^2 (M) = \frac{1}{2\pi^2}\int^{\infty}_0 k^2 P(k) \tilde{W}^2(kR) {\rm d}k,
\end{equation}
where $\tilde{W}(kR)$ is the Fourier transform of the spherical top-hat window function,
\begin{equation}
\tilde{W}(kR) = \frac{3\sin (kR) - (kR) \cos (kR)}{(kR)^3}.
\end{equation}  
The radius $R$ is related to the mass $M$ by
\begin{equation}
R = \left( \frac{M}{\frac{4}{3}\pi \rho_m} \right)^{\frac{1}{3}}.
\end{equation}  
where $\rho_m = \frac{3 H_0^2}{8 \pi G} \times \Omega_{\rm M} = 2.78 \times 10^{11} \,\Omega_{\rm M} \,h^2 \,M_{\sun} \,{\rm Mpc}^{-3}$ is the mean density of the Universe at $z = 0$.
$P(k)$ is the power spectrum of the density perturbations,
\begin{equation}
P(k) = k^{n_{\rm s}} T^2 (k),
\end{equation}  
The amplitude of $P(k)$ is fixed to match $\sigma_8$, the density variance in a sphere of radius 8 $h^{-1} $ Mpc.
$T(k)$ is the CDM transfer function given by \citet{Eisenstein},
\begin{eqnarray}
T (k) &=& \frac{\ln [2e + 1.8q(k)]}{\ln [2e + 1.8q(k)] + \left[14.2 + \frac{731}{1+62.5q(k)} \right]q^2(k)}, \\
q(k) &=& \left(\frac{k}{h\, {\rm Mpc^{-1}}}\right) \frac{1}{\Gamma (k)}
\end{eqnarray}
where $\Gamma (k)$ is the CDM shape parameter given by \citet{Eisenstein}
\begin{equation}
\Gamma (k) = \Omega_{\rm M}h \left[\alpha_{\Gamma} + \frac{1-\alpha_{\Gamma}}{(1+0.43ks)^4} \right],
\end{equation}  
where 
\begin{eqnarray}
\alpha_{\Gamma} &=&  1 - 0.328 \ln(431\Omega_{\rm M}h^2)\frac{\Omega_{\rm b}}{\Omega_{\rm M}}  \nonumber \\
&+& 0.38 \ln(22.3\Omega_{\rm M}h^2) \left(\frac{\Omega_{\rm b}}{\Omega_{\rm M}} \right)^2\\
s &=& \frac{44.5\ln\left(\frac{9.83}{\Omega_{\rm M}h^2} \right)}{\sqrt{1+10(\Omega_{\rm b}h^2)^{\frac{3}{4}}}} {\rm Mpc}.
\label{s}
\end{eqnarray}
Given our two models for the redshift distribution of DOGs, and our measured values of $b_{\rm DOG}$, we  estimate the DM halo mass in which they reside based on Equation (\ref{d_M}) -- (\ref{s}).
We define the mean halo mass $\langle M_{\mathrm{h}} \rangle$ as the mass which satisfies the following equation,
\begin{equation}
b_{\mathrm{DOG}} = b(\langle M_{\mathrm{h}} \rangle, z),
\end{equation}
where $b(M, z)$ is the halo bias from Equation (\ref{d_M}). 
As a result, we obtained $\log [\langle M_{\mathrm{h}} \rangle / (h^{-1} M_{\sun})]$ $\sim$ 13.57$_{-0.55}^{+0.50}$ and 13.65$_{-0.52}^{+0.45}$ in Case 1 and 2, respectively.
We similarly found that the bias values of IR-faint DOGs from \cite{Brodwin} show that they reside in DM halos with $8.7\times 10^{12} M_{\sun}$, roughly consistent with the value of \cite{Brodwin}.
However, our IR-bright DOG sample resides in significantly more massive DM halos, as shown in Figure \ref{bias_comp}(a).
Comparing the bias factor and DM halo masses of the IR-bright DOGs with other populations, we now discuss the possibility of an evolutionary connection between them.
\citet{Dey} suggested that SMGs, DOGs and quasars represent phases in the evolution of a gas-rich
merger at high redshift in the context of the major merger scenario (\citealt{Sanders,Hopkins_06}, see also \citealt{Hickox}).
Hydrodynamic simulations conducted by \citet{Narayanan} suggest that DOGs represent a transition phase in the evolutionary sequence of galaxy mergers; the sequence progresses from SMGs to DOGs to quasars to
elliptical galaxies.
In the case of IR-faint DOGs, we confirmed that they reside in DM halos with similar mass as SMGs and SDSS quasars at similar redshift, which supports the evolutionary link between them as suggested by previous works.
On the other hand, our result argues against an evolutionary link between IR-bright DOGs and other high-redshift populations.
We found that IR-bright DOGs reside in several times heavier DM halos than do IR-faint DOGs and SMGs.
This significant difference of DM halo mass cannot be explained as an evolutionary sequence involving major mergers.
Therefore, IR-bright DOGs differ from other high-redshift populations and there is no evolutionary link between them. 

However, SDSS quasars and our IR-bright DOGs reside in DM halos with similar mass and thus they are likely to be on similar evolutionary track for halo bias factor (see Figure \ref{bias_comp}a), which indicates that they could have an evolutionary link.
We note that the bolometric luminosities of the SDSS quasars plotted in Figure \ref{bias_comp} lies in the range $\log \,(L_{\rm bol}/L_{\sun})$ = 11.6 -- 13.5 \citep{Shen,Ross}.
We estimated the bolometric luminosity for the galaxies in our sample; we first estimate the IR luminosity of each object based on the Monte Carlo method given the redshift distribution in the same manner as \cite{Toba} and then converted it to bolometric luminosity assuming the bolometric correction, where we adopted $L_{\rm bol} = 1.4 \times L_{\rm IR}$ \citep{Fan}.
The estimated bolometric luminosity of IR-bright DOGs is $\log \,(L_{\rm bol}/L_{\sun})$ = 13.6 $\pm$ 0.2 and 13.0 $\pm $ 0.3 for Case 1 and 2, respectively.
We also note that \cite{Shen} examined the clustering of the 10\% most luminous quasars in their sample at $z$ = 0.4 -- 2.5.
The resultant bias factor is $b \sim$ 5.2, larger than the remaining 90\% of quasars, and roughly consistent with our IR-bright DOG sample.
Their bolometric luminosity is $\log \,(L_{\rm bol}/L_{\sun})$ $\sim$ 13.3 while their number density estimated from the LF in \cite{Richards} is roughly $5.0 \times 10^{-7}$ Mpc$^{-3}$ dex$^{-1}$ at $z$ = 2.01 and $1.3 \times 10^{-7}$ at $z$ = 1.25, which are also in good agreement with those in \cite{Toba}. 
Therefore, IR-bright DOGs may well have an evolutionary link with the most luminous SDSS quasars at $z \sim$ 1 -- 2.

The lines in Figure \ref{bias_comp}(b) show the evolution of bias from the observed epoch ($z \sim 1-2$) to the present day ($z = 0$) assuming that DOGs passively evolve into their lower redshift counterparts with the bias evolution given by \cite{Fry} (see also \citealt{White}).
In both Case 1 and Case 2, we found that the estimated DM halo mass of IR-bright DOGs is several times heavier than that of IR-faint DOGs and they follow different evolutionary tracks. 
This also indicates that IR-bright DOGs are a different population from IR-faint DOGs and thus there is no evolutionary link between them.

Figure \ref{bias_comp}(b) also indicates that the bias of IR-faint DOGs will evolve to $b$ = $2.18_{-0.18}^{+0.26}$ at $z$ = 0 while that of IR-bright DOGs will evolve to $b$ = $3.10_{-0.62}^{+0.82}$ and $2.62_{-0.49}^{+0.62}$ for Case 1 and 2, respectively.
This implies that the IR-bright DOGs will reside in very massive DM halos with $3.2_{-1.5}^{+2.6} \times 10^{14} \,h^{-1} M_{\sun}$ and $2.0_{-1.0}^{+1.6} \times 10^{14} \, h^{-1} M_{\sun}$ at $z = 0$ in Case 1 and 2, respectively, roughly comparable to the mass of present-day moderately massive galaxy clusters such as the Virgo cluster \citep{Karachentsev}.
Therefore, we conclude that IR-bright DOGs may trace the mass assembly of stars and SMBHs through the heavy AGN/SF activity within massive DM halos in the high-redshift ($z \sim$ 1 -- 3) Universe, and that IR-bright DOGs could be progenitors of moderately massive galaxy clusters in the present epoch.

%-----------------
% Minimum halo mass and duty cycle
%-----------------
\subsection{Minimum halo mass and duty cycle}
Here, we discuss the duty cycle of the phase in which a galaxy appears as a DOG.
First, we estimate the minimum halo mass ($M_{\mathrm{h,min}}$) by comparing the bias of the DOGs with the effective bias of DM halos with this minimum halo mass.
The effective bias is defined as
\begin{equation}
b_{\mathrm{eff}} = \frac{ \int_{M_{\mathrm{h,min}}}^{\infty} b(M) \frac{\mathrm{d}n}{\mathrm{d}M} (M)~\mathrm{d} M } { \int_{M_{\mathrm{h,min}}}^{\infty} \frac{\mathrm{d}n}{\mathrm{d}M} (M)~\mathrm{d} M },
\end{equation}
where $\frac{\mathrm{d}n}{\mathrm{d}M} (M)~\mathrm{d} M$ is the number density of haloes 
with mass between $M$ and $M+\mathrm{d}M$, and $b(M)$ is given by Equation (\ref{d_M}).
In this work, we adopt the halo mass function $\frac{\mathrm{d}n}{\mathrm{d}M} (M)$ given by \citet{Sheth};
\begin{equation}
\frac{\mathrm{d}n}{\mathrm{d}M} (M) = \frac {\rho_m} {M^{2}} f(\nu) \left | \frac {\mathrm{d}\ln \nu} {\mathrm{d}\ln M} \right |, 
\end{equation}
where $\nu = \frac{\delta_c}{\sigma (M)} = \frac{1.686}{\sigma (M)}$ and $\sigma (M)$ was defined in Equation (\ref{sigma_M}).
$f ({\nu})$ is given by
\begin{equation}
f(\nu) = 2A \left ( 1 + \frac {1} {{\nu'}^{2q}}  \right ) \left ( \frac {{\nu'}^2} {2\pi} \right )^{1/2} \exp \left ( - \frac {{\nu'}^2} {2} \right ),
\end{equation}
where $\nu' = \sqrt{a} \nu$, and $a=0.707$, $q=0.3$, and A=0.322.
We obtained $\log [\langle M_{\mathrm{h, min}} \rangle / (h^{-1} M_{\sun})]$ = 13.40$_{-0.59}^{+0.53}$ and 13.41$_{-0.56}^{+0.49}$ in Case 1 and 2, respectively (see Table \ref{Table}).

We then constrain the duty cycle of DOGs following previous works on quasars \citep{Martini,Haiman,Eftekharzadeh}.
Given this result, we estimated the DOG number density and the predicted number density of DM halos with mass above the minimum halo mass we estimate.
Here, we assume that all DOGs reside in DM halos with mass above $M_{\mathrm{h,min}}$ and that each halo with mass above $M_{\mathrm{h,min}}$ hosts at most one DOG at a time.
We define the duty cycle as follows:
\begin{equation}
f_{\mathrm{duty}} = \frac{ \int_{L_{\mathrm{min}}}^{L_{\mathrm{max}}} \Phi (L)~\mathrm{d} L } { \int_{M_{\mathrm{h,min}}}^{\infty} \frac{\mathrm{d}n}{\mathrm{d}M} (M)~\mathrm{d} M },
\end{equation}
where $\Phi (L)$ is the DOG luminosity function (LF).
We recalculated the LF and number density of DOGs using the DOG sample in \cite{Toba} under the Planck cosmology.
One caution here is that the derived number density of IR-bright DOGs should be considered as a lower limit because the DOG sample could miss some IR-bright DOGs due to limitations in the version of the HSC pipeline that was used for that analysis.
In addition, the estimate of the duty cycle is sensitive to the uncertainty of the derived DM halo mass.
We find $f_{\rm duty}$ $>$ 0.013 and 0.007 for Case 1 and 2, respectively.

If we assume that the lifetime of DM halos in which DOGs reside is approximately the age of the Universe at its redshift, the lifetime of the DOG phase is $>$ 0.013 $\times$ 3.27 Gyr $\sim$ 41 Myr and $>$ 0.007 $\times$ 5.13 Gyr $\sim$ 37 Myr for Case 1 and 2, respectively.
\citet{Narayanan} combined N-body/smoothed particle hydrodynamic simulations with 3D polychromatic dust radiative transfer models, and investigated the IR to optical flux ratio in major mergers as a function of cosmic time.
They found that the phase during which $F_{\rm MIR}$/$F_{\rm opt}$ $>$ 1000 (thus matching the definition of a DOG) was only 10-30 Myr, somewhat shorter than what we had found.  
However, the system that \cite{Narayanan} simulated would have a MIR flux of less than 1 mJy if placed at $z \sim$ 2, and thus this object would be an IR-faint DOG.  
In addition, this simulation is for the coalescence of two galaxies with total mass of $3.4 \times 10^{12} M_{\sun}$.
Therefore, given that (1) the IR-bright DOGs reside in more massive DM halos than do IR-faint DOGs and (2) major mergers with more massive galaxies form DOGs with a longer duty cycle \citep{Narayanan}, the fact that the duty cycle we had found is longer than that obtained from the simulation would be consistent.
The IR-bright DOGs are expected to result from mergers with more massive galaxies ($> 10^{13} \,M_{\sun}$) than are IR-faint DOGs.

%======================
%      SUMMARY
%======================
\section{Summary}
\label{Summary}
Using the latest HSC-wide survey data and the {\it WISE} MIR all-sky survey data, we performed a search for IR-bright DOGs over $\sim$ 125 deg$^2$.
We cross-matched the HSC $i$-band photometry with the {\it WISE} 22 $\micron$ catalog, and adopted the DOGs color selection ($i$ - [22] $>$ 7.0).
We identified 4,367 IR-bright DOGs with flux $>$ 1.0 mJy at 22 $\micron$. 
After extracting a uniform subsample of 1,411 galaxies, we measured their angular autocorrelation function (ACF), from which we determined the correlation length, and dark matter halo mass in which they reside.
We consider possible two cases for their redshift distribution.
The main results are as follows:
\begin{enumerate}
\item The IR-bright DOG ACF follows a power-law, $\omega (\theta)$ = (0.010 $\pm$ 0.003) $(\theta/{\rm deg.})^{-0.9}$. The correlation amplitude of IR-bright DOGs is larger than that of IR-faint DOGs, which confirms that the clustering of DOGs depends on MIR flux \citep{Brodwin}.
\item Assuming that the redshift distribution for our DOG sample is Gaussian, with mean and sigma $z$ = 1.99 $\pm$ 0.45 (Case 1) and $z$ = 1.19 $\pm$ 0.30 (Case 2), the correlation length of IR-bright DOGs is $r_0$ = 12.0 $\pm$ 2.0 and 10.3 $\pm$ 1.7 $h^{-1}$ Mpc for Case 1 and 2, respectively.
\item Given the correlation length, we found that their bias factor is $b_{\rm DOG}$ = 5.99 $\pm$ 0.96 and 3.88 $\pm$ 0.62 for Case 1 and 2, respectively and they reside in massive dark matter halos with masses $\log [\langle M_{\mathrm{h}} \rangle / (h^{-1} M_{\sun})]$ $\sim$ 13.57$_{-0.55}^{+0.50}$ and 13.65$_{-0.52}^{+0.45}$ in Case 1 and 2, respectively.
\item The derived bias factor and dark matter halo mass of IR-bright DOGs are larger than those of other populations such as IR-faint DOGs and SMGs, while they are consistent with that of the SDSS quasars (particularly with $\log L_{\rm bol}$ $\sim$ 13.3 $L_{\sun}$). This indicates that IR-bright DOGs differ from IR-faint DOGs and SMGs, but may have an evolutionary link with the SDSS quasars.
\item Their duty cycle estimated by the ratio of number densities of IR-bright DOGs and DM halos in which they reside is at least 0.013 and 0.007 in Case 1 and 2, respectively.
\end{enumerate}

The current sample is based on $\sim$ 125 deg$^2$ of imaging data. 
The HSC survey will cover more than 10 times as much sky, 1400 deg$^2$ when it is complete, allowing the identification of more than 10,000 IR-bright DOGs.  
This sample will be large enough to carry out a variety of statistical analyses to understand the physical nature of these objects (e.g., executing the clustering analysis on larger scale, doing a halo occupation distribution analysis and investigating the clustering strength as a function of flux).
We are also searching for DOGs in the deep fields (27 deg$^2$) of the HSC survey: going a magnitude deeper in optical filters will allow identification of even redder sources than are explored in this paper.

\acknowledgments
The authors gratefully acknowledge the anonymous referee for a careful reading of the manuscript and very helpful comments.
We thank Prof. Mark Brodwin for providing the ACF data for IR-faint DOGs.
The Hyper Suprime-Cam (HSC) collaboration includes the astronomical communities of Japan and Taiwan, and Princeton University. The HSC instrumentation and software were developed by the National Astronomical Observatory of Japan (NAOJ), the Kavli Institute for the Physics and Mathematics of the Universe (Kavli IPMU), the University of Tokyo, the High Energy Accelerator Research Organization (KEK), the Academia Sinica Institute for Astronomy and Astrophysics in Taiwan (ASIAA), and Princeton University.  Funding was contributed by the FIRST program from Japanese Cabinet Office, the Ministry of Education, Culture, Sports, Science and Technology (MEXT), the Japan Society for the Promotion of Science (JSPS),  Japan Science and Technology Agency (JST),  the Toray Science  Foundation, NAOJ, Kavli IPMU, KEK, ASIAA, and Princeton University.
This paper makes use of software developed for the Large Synoptic Survey Telescope. We thank the LSST Project for making their code available as free software at http://dm.lsstcorp.org.
The Pan-STARRS1 Surveys (PS1) have been made possible through contributions of the Institute for Astronomy, the University of Hawaii, the Pan-STARRS Project Office, the Max-Planck Society and its participating institutes, the Max Planck Institute for Astronomy, Heidelberg and the Max Planck Institute for Extraterrestrial Physics, Garching, The Johns Hopkins University, Durham University, the University of Edinburgh, Queen's University Belfast, the Harvard-Smithsonian Center for Astrophysics, the Las Cumbres Observatory Global Telescope Network Incorporated, the National Central University of Taiwan, the Space Telescope Science Institute, the National Aeronautics and Space Administration under Grant No. NNX08AR22G issued through the Planetary Science Division of the NASA Science Mission Directorate, the National Science Foundation under Grant No. AST-1238877, the University of Maryland, and Eotvos Lorand University (ELTE).
This research was supported by World Premier International Research Center Initiative (WPI), MEXT, Japan and by MEXT as ``Priority Issue on Post-K computer'' (Elucidation of the Fundamental Laws and Evolution of the Universe) and JICFuS.
TN is financially supported by JSPS KAKENHI (grant numbers 25707010, 16H01101, and 16H03958) and also by the JCG-S Scholarship Foundation. 
Y.Terashima is also financially supported by JSPS KAKENHI (grant number 15H02070).

\listofchanges


\begin{thebibliography}{}

\bibitem[Abazajian et al. (2004)]  {Abazajian}
Abazajian, K., Adelman-McCarthy, J. K., Ag\"ueros, M. A., et al.\ 2004, \aj, 128, 502
\bibitem[Alam et al. (2015)]  {Alam}
Alam, S., Albareti, F. D., Allende Prieto, C., et al.\ 2015, \apjs, 219
\bibitem[Allevato et al. (2014)]  {Allevato}
Allevato, V., Finoguenov, A., Civano, F., et al.\ 2014, \apj, 796
\bibitem[Arnaboldi et al. (2007)]  {Arnaboldi}
Arnaboldi, M., Neeser, M. J., Parker, L. C., et al.\ 2007, Msngr, 127, 28
\bibitem[Axelrod et al. (2010)]  {Axelrod}
Axelrod, T., Kantor, J., Lupton, R. H., \& Pierfederici, F. 2010, Proc. SPIE, 7740, 15
\bibitem[Blain et al. (2004)]  {Blain}
Blain, A. W., Chapman, S. C., Smail, I., \& Ivison, R. 2004, \apj, 611, 725
\bibitem[Blain et al. (2002)]  {Blain_02}
Blain, A. W., Smail, I., Ivison, R. J., Kneib, J.-P., \& Frayer, D. T. 2002, Physics Reports, 369, 111
\bibitem[Bouwens et al. (2011)]  {Bouwens}
Bouwens, R. J., Illingworth, G. D., Labbe, I., et al.\ 2011, Nature, 469, 504
\bibitem[Brand et al. (2006)]  {Brand}
Brand, K., Dey, A., Weedman, D., et al.\ 2006, \apj, 644, 143
\bibitem[Brodwin et al. (2008)]  {Brodwin}
Brodwin, M., Dey, A., Brown, M. J. I., et al.\ 2008, \apjl, 687, L65
\bibitem[Bruzual \& Charlot (2003)]  {Bruzual}
Bruzual, G., \& Charlot, S. 2003, \mnras, 344, 1000
\bibitem[Bussmann et al. (2012)]  {Bussmann_12}
Bussmann, R. S., Dey, A., Armus, L., et al.\ 2012, \apj, 744, 150
\bibitem[Bussmann et al. (2009)]  {Bussmann_09}
Bussmann, R. S., Dey, A., Borys, C., et al.\ 2009, \apj, 705, 184
\bibitem[Bussmann (2011)]  {Bussmann_11}
Bussmann, R. S., Dey, A., Lotz, J., et al.\ 2011, \apj, 733, 21
\bibitem[Cutri et al. (2014)]{Cutri}
Cutri, R. M. 2014, yCat, 2328, 0
\bibitem[Desai et al. (2009)]  {Desai}
Desai, V., Soifer, B. T., Dey, A., et al.\ 2009, \apj, 700, 1190
\bibitem[Dey et al. (2008)]  {Dey}
Dey, A., Soifer, B. T., Desai, V., et al.\ 2008, \apj, 677, 943
\bibitem[Donoso et al. (2014)]  {Donoso}
Donoso, E., Yan, L., Stern, D., \& Assef, R. J. 2014, \apj, 789, 44
\bibitem[Driver et al. (2011)]  {Driver_11}
Driver, S. P., Hill, D. T., Kelvin, L. S., et al.\ 2011, \mnras, 413, 971
\bibitem[Driver et al. (2009)]  {Driver_09}
Driver, S. P., Norberg, P., Baldry, I. K., et al.\ 2009, A\&G, 50, 12
\bibitem[Eftekharzadeh et al. (2015)]  {Eftekharzadeh}
Eftekharzadeh, S., Myers, A. D., White, M., et al.\ 2015, \mnras, 453, 2779
\bibitem[Eisenstein \& Hu (1998)]  {Eisenstein}
Eisenstein, D. J., \& Hu, W. 1998, \apj, 496, 605
\bibitem[Fan et al. (2016)]  {Fan}
Fan, L., Han, Y., Nikutta, R., Drouart, G., \& Knudsen, K. K. 2016, \apj, 823, 107
\bibitem[Farrah et al. (2006)]  {Farrah}
Farrah, D., Lonsdale, C. J., Borys, C., et al.\ 2006, \apjl, 643, L139
\bibitem[Fry (1996)]  {Fry}
Fry, J. N. 1996, \apjl, 461, L65
\bibitem[Goto et al. (2011)]  {Goto}
Goto, T., Arnouts, S., Inami, H., et al.\ 2011, \mnras, 410, 573
\bibitem[Groth \& Peebles (1977)]  {Groth}
Groth, E. J., \& Peebles, P. J. E. 1977, \apj, 217, 385
\bibitem[Guglielmo et al. (2015)]  {Guglielmo}
Guglielmo, V., Poggianti, B. M., Moretti, A., et al.\ 2015, \mnras, 450, 2749
\bibitem[Haiman \& Hui (2001)]  {Haiman}
Haiman, Z., \& Hui, L. 2001, \apj, 547, 27
\bibitem[Hewett (1982)]  {Hewett}
Hewett, P. C. 1982, \mnras, 201, 867
\bibitem[Hickox et al. (2009)]  {Hickox}
Hickox, R. C., Jones, C., Forman, W. R., et al.\ 2009, \apj, 696, 891
\bibitem[Hickox et al. (2012)]  {Hickox_12}
Hickox, R. C., Wardlow, J. L., Smail, I., et al.\ 2012, \mnras, 421, 284
\bibitem[Hopkins et al. (2006)]{Hopkins_06}
Hopkins, P. F., Hernquist, L., Cox, T. J., Di Matteo, T., Robertson, B., \& Springel, V. 2006, \apjs, 163, 1
\bibitem[Houck et al. (2005)]  {Houck}
Houck, J. R., Soifer, B. T., Weedman, D., et al.\ 2005, \apjl, 622, L105
\bibitem[Guglielmo et al. (2015)]  {Guglielmo}
Guglielmo, V., Poggianti, B. M., Moretti, A., Fritz, J., Calvi, R., Vulcani, B., Fasano, G., \& Paccagnella, A. 2015, \mnras, 450, 2749
\bibitem[Ikeda et al. (2015)]  {Ikeda}
Ikeda, H., Nagao, T., Taniguchi, Y., et al.\ 2015, \apj, 809
\bibitem[Ivezi\'c et al. (2008)]{Ivezic}
Ivezi\'c, \v{Z}., Tyson, J. A., Abel, B., et al. 2008, arXiv:0805.2366
\bibitem[Jannuzi \& Dey (1999)]{Jannuzi}
Jannuzi, B. T., \& Dey, A. 1999, in ASP Conference Series, Vol. 191, Photometric Redshifts and the Detection of High Redshift Galaxies, ed. R. Weymann et al.\ (San Francisco, CA: ASP), 111
\bibitem[Juri\'c et al. (2015)]{Juric}
Juri\'c, M., Kantor, J., Lim, K., et al.\ 2015, arXiv:1512.07914
\bibitem[Karachentsev et al. (2014)]{Karachentsev}
Karachentsev, I. D., Tully, R. B., Wu, P.-F., Shaya, E. J., \& Dolphin, A. E. 2014, \apj, 782
\bibitem[Landy \& Szalay (1993)]  {Landy}
Landy, S. D., \& Szalay, A. S. 1993, \apj, 412, 64
\bibitem[Limber  (1954)]  {Limber}
Limber, D. N. 1954, \apj, 119, 655
\bibitem[Lupton et al. (2001)]{Lupton}
Lupton, R. H. et al., 2001, SDSS Image Processing {II}: The \emph{Photo} Pipelines
http://www.astro.princeton.edu/~rhl/photo-lite.pdf
\bibitem[Ling et al. (1986)]{Ling}
Ling, E. N., Barrow, J. D., \& Frenk, C. S. 1986, \mnras, 223, 21P
\bibitem[Madau \& Dickinson (2014)]  {Madau}
Madau, P., \& Dickinson, M. 2014, \araa, 52, 415
\bibitem[Magliocchetti et al. (2008)]  {Magliocchetti}
Magliocchetti, M., Cirasuolo, M., McLure, R. J., et al.\ 2008, \mnras, 383, 1131
\bibitem[Magnier et al. (2013)]  {Magnier}
Magnier, E. A., Schlafly, E., Finkbeiner, D., et al.\ 2013, \apjs, 205, 20
\bibitem[Magorrian et al. (1998)]{Magorrian}
Magorrian, J., Tremaine, S., Richstone, D., et al.\ 1998, \aj, 115, 2285
\bibitem[Marconi \& Hunt (2003)]{Marconi}
Marconi, A., \& Hunt, L. K. 2003, \apjl, 589, L21
\bibitem[Martini \& Weinberg (2001)]  {Martini}
Martini, P., \& Weinberg, D. H. 2001, \apj, 547, 12
\bibitem[Melbourne et al. (2012)]  {Melbourne}
Melbourne, J., Soifer, B. T., Desai, V., et al.\ 2012, \aj, 143, 125
\bibitem[Miyazaki et al. (2012)]{Miyazaki}
Miyazaki, S., Komiyama, Y., Nakaya, H., et al.\ 2012, Proc. SPIE, 8446, 0
\bibitem[Myers et al. (2006)]  {Myers}
Myers, A. D., Brunner, R. J., Richards, G. T., et al.\ 2006, \apj, 638, 622
\bibitem[Narayanan et al. (2010)]{Narayanan}
Narayanan, D., Dey, A., Hayward, C. C., et al.\ 2010, \mnras, 407, 1701
\bibitem[Peebles (1980)]  {Peebles}
Peebles, P. J. E. 1980, The Large-scale Structure of the Universe (Princeton,
NJ: Princeton Univ. Press)
\bibitem[Pope et al. (2008)]  {Pope}
Pope, A., Bussmann, R. S., Dey, A., et al. 2008, \apj, 689, 127
\bibitem[Planck Collaboration (2014)]  {Planck}
Planck Collaboration XVI, 2014, \aap, 571, A16
\bibitem[Richards et al. (2006)]  {Richards}
Richards, G. T., Strauss, M. A., Fan, X., et al.\ 2006, \aj, 131, 2766
\bibitem[Roche \& Eales (1999)]  {Roche}
Roche, N., \& Eales, S. A. 1999, \mnras, 307, 703
\bibitem[Ross et al. (2009)]  {Ross}
Ross, N. P., Shen, Y., Strauss, M. A., et al.\ 2009, \apj, 697, 1634
\bibitem[Sanders et al. (1988)]  {Sanders}
Sanders, D. B., Soifer, B. T., Elias, J. H., et al.\ 1988, \apj, 325, 74
\bibitem[Schlafly et al. (2012)]  {Schlafly}
Schlafly, E. F., Finkbeiner, D. P., Juri\'c, M., et al.\ 2012, \apj, 756, 158
\bibitem[Shen et al. (2009)]  {Shen}
Shen, Y., Strauss, M. A., Ross, N. P., et al.\ 2009, \apj, 697, 1656
\bibitem[Sheth et al. (2001)]  {Sheth}
Sheth, R. K., Mo, H. J., \& Tormen, G. 2001, \mnras, 323, 1
\bibitem[Tanaka (2015)]  {Tanaka}
Tanaka, M. 2015, \apj, 801
\bibitem[Toba \& Nagao (2016)]  {Toba_16}
Toba, Y., \& Nagao, T. 2016, \apj, 820, 46
\bibitem[Toba et al. (2015)]  {Toba}
Toba, Y., Nagao, T., Strauss, M. A., et al.\ 2015, \pasj, 67, 86
\bibitem[Taylor et al. (2005)]{Taylor}
Taylor, M. B., Britton, M., \& Ebert, R., 2005, in ASP Conf. Ser. 347, Astronomical Data Analysis Software and Systems XIV, ed. P. Shopbell, M. Britton, \& R. Ebert (San Francisco, CA: ASP), 29
\bibitem[Tonry et al. (2012)]  {Tonry}
Tonry, J. L., Stubbs, C. W., Lykke, K. R., et al.\ 2012, \apj, 750, 99
\bibitem[White et al. (2007)]  {White}
White, M., Zheng, Z., Brown, M. J. I., Dey, A., \& Jannuzi, B. T. 2007, \apjl, 655, L69
\bibitem[Williams et al. (2011)]  {Williams}
Williams, C. C., Giavalisco, M., Porciani, C., et al. 2011, \apj, 733
\bibitem[Wright et al. (2010)]  {Wright}
Wright, E. L., Eisenhardt, P. R. M., Mainzer, A. K., et al.\ 2010, \aj, 140, 1868
\bibitem[Yan et al. (2013)]  {Yan}
Yan, L., Donoso, E., Tsai, C.-W., et al.\ 2013, \aj, 145, 55
\bibitem[York et al. (2000)]  {York}
York, D. G., Adelman, J., Anderson, J. E., Jr., et al.\ 2000, \aj, 120, 1579
\end{thebibliography}
\end{document}